\documentclass{IEEEtran-final}
\pdfoutput=1

\usepackage{amsfonts}
\usepackage{xcolor}
\usepackage{hyperref}
\usepackage{subcaption} 
\usepackage{amsmath}
\usepackage{amssymb}
\usepackage{mathtools}
\usepackage{multirow}
\usepackage{enumitem}
\usepackage{graphicx}
\usepackage[export]{adjustbox}
\usepackage{cite}
\usepackage{balance}
\usepackage{xspace}
\usepackage[normalem]{ulem} 

\usepackage{arydshln} 

\usepackage{algorithm} 
\usepackage[noend]{algpseudocode} 
\usepackage{etoolbox}

\makeatletter
\newcommand*{\algline}[1][\algorithmicindent]{%
    \hspace*{.2em}%
    {\color{black!20}\vrule}%
    \hspace*{\dimexpr#1-.2em-.4pt}%
}

\renewcommand{\ALG@beginalgorithmic}{\offinterlineskip}

\newcount\ALG@printindent@tempcnta
\def\ALG@printindent{%
    \ifnum \theALG@nested > 0
        \ifx\ALG@text\ALG@x@notext 
        \else
            \unskip
            \ALG@printindent@tempcnta=1
            \loop
                \algline[\csname ALG@ind@\the\ALG@printindent@tempcnta\endcsname]%
                \advance \ALG@printindent@tempcnta 1
                \ifnum \ALG@printindent@tempcnta<\numexpr\theALG@nested+1\relax
            \repeat
        \fi
    \fi
}

\patchcmd{\ALG@doentity}{\noindent\hskip\ALG@tlm}{\ALG@printindent}{}{\errmessage{failed to patch}}
\makeatother

\algrenewcommand\algorithmicend{\strut\textbf{end}}
\algrenewcommand\algorithmicdo{\strut\textbf{do}}
\algrenewcommand\algorithmicwhile{\strut\textbf{while}}
\algrenewcommand\algorithmicfor{\strut\textbf{for}}
\algrenewcommand\algorithmicforall{\strut\textbf{for all}}
\algrenewcommand\algorithmicloop{\strut\textbf{loop}}
\algrenewcommand\algorithmicrepeat{\strut\textbf{repeat}}
\algrenewcommand\algorithmicuntil{\strut\textbf{until}}
\algrenewcommand\algorithmicprocedure{\strut\textbf{procedure}}
\algrenewcommand\algorithmicfunction{\strut\textbf{function}}
\algrenewcommand\algorithmicif{\strut\textbf{if}}
\algrenewcommand\algorithmicthen{\strut\textbf{then}}
\algrenewcommand\algorithmicelse{\strut\textbf{else}}

\algrenewcommand\algorithmicrequire{\strut\textbf{Input:}}
\algrenewcommand\algorithmicensure{\strut\textbf{Output:}}

\let\oldState\State
\renewcommand{\State}{\oldState\strut}

\newcommand{\xrep}{\vc x^{\text{rep}}}
\newcommand{\mat}[1]{{\boldsymbol{#1}}}
\newcommand{\verts}{\mathcal{V}}
\newcommand{\XX}{\mathcal{X}}
\newcommand{\YY}{\mathcal{Y}}

\newcommand{\normals}{\mathcal{N}}
\newcommand{\defD}{\mathcal{D}}

\newcommand{\grp}{\Aut(\BW)}
\newcommand{\grpG}{\mathcal{G}}
\newcommand{\grpS}{\mathcal{S}}
\newcommand{\grpP}{\mathcal{P}}

\newcommand{\setU}{\mathcal{U}}
\newcommand{\setT}{\mathcal{T}}

\newcommand{\setM}{\mathcal{M}}

\newcommand{\defeq}{=}

\newcommand\tstrut{\rule{0pt}{2.4ex}}
\newcommand\bstrut{\rule[-1.2ex]{0pt}{0pt}}

\newcommand{\CommentFont}[1]{{\footnotesize\textcolor{gray!80!black}{#1}}}
\newcommand{\CommentLeader}{{\CommentFont{%
}}}
\newcommand{\CommentInlineText}[1]{{\CommentFont{%
    \CommentLeader\ignorespaces #1\unskip%
}}}

\makeatletter
\newcommand{\CommentLineEmpty}[2][0]{%
    \addtocounter{ALG@nested}{-#1}%
    \Statex\strut\ALG@printindent\relax\CommentInlineText{#2}%
    \addtocounter{ALG@nested}{#1}%
}
\makeatother

\newcommand{\proc}[1]{{\scshape\small #1}}
\newcommand{\class}[1]{{``#1''}}

\newcommand{\lib}[1]{{\em #1}}

\newcommand{\eqnref}[1]{\eqref{#1}}

\newcommand{\Z}{{\mathbb{Z}}}
\newcommand{\vc}[1]{{\boldsymbol{#1}}}
\newcommand{\E}{{\mathbb{E}}}

\newcommand{\BW}{\Lambda_{16}}

\DeclareMathOperator{\Aut}{Aut}
\DeclareMathOperator{\Stab}{Stab}

\DeclareMathOperator*{\argmin}{arg\,min}
\DeclareMathOperator{\var}{var}
\DeclareMathOperator{\varh}{\widehat{var}}
\DeclareMathOperator{\tr}{tr}

\newenvironment{alg}[1][tbp]{
    \begin{algorithm}[#1]
}{
    \end{algorithm}
}

\IfElseFinal{%
    \year=2023 \month=01 \day=15
    \markboth{Preprint, \today}{Preprint, \today}%
}{}

\begin{document}

\title{The Voronoi Region of the Barnes--Wall Lattice $\Lambda_{16}$}
\author{%
    Daniel Pook-Kolb,
    Erik Agrell, \IEEEmembership{Fellow, IEEE}, and
    Bruce Allen, \IEEEmembership{Member, IEEE}
    \IfElseFinal{%
    \thanks{%
        This work has been submitted to the IEEE for possible publication.
        Copyright may be transferred without notice, after which this version may no
        longer be accessible.
    }}{}%
    \thanks{%
        D.~Pook-Kolb and B.~Allen are with the
        Max Planck Institute for Gravitational Physics (Albert Einstein Institute),
        30167 Hannover, Germany, and
        Leibniz Universit\"at Hannover
        (e-mail: daniel.pook.kolb@aei.mpg.de and bruce.allen@aei.mpg.de).
    }
    \thanks{%
        E.~Agrell is with the
        Department of Electrical Engineering,
        Chalmers University of Technology, 41296 Gothenburg, Sweden
        (e-mail: agrell@chalmers.se).
    }
}

\maketitle

\begin{abstract}
    We give a detailed description of the Voronoi region of the
    Barnes--Wall lattice $\BW$, including its vertices, relevant vectors,
    and symmetry group.
    The exact value of its quantizer constant is calculated,
    which was previously only known approximately.
To verify the result, we estimate the same constant numerically and propose a new
very simple method to quantify the variance of such estimates, which is
far
more accurate than the commonly used jackknife estimator.
\end{abstract}


\section{Introduction}
\label{sec:intro}

\IEEEPARstart{I}{n} 1959, E.~S.\ Barnes and G.~E.\ Wall introduced a family of lattices in
dimensions $4,8,16,\ldots$ based on Abelian groups \cite{barnes59}. In
dimensions $4$ and $8$, the proposed construction reproduced known lattices,
which are nowadays denoted as $D_4$ and $E_8$, respectively, whereas
previously unknown lattices were obtained in dimensions $16$ and up.
Alternative constructions and further properties of the \emph{Barnes--Wall
(BW) lattices} were investigated in \cite{forney88pt1, forney88pt2, hahn90}.

The BW lattices are remarkably good in three of the standard figures of merit
for lattices: \emph{packing,} \emph{kissing,} and \emph{quantization.} In
fact, they are known or conjectured to be optimal in all three figures of
merit in dimensions $n=4$, $8$, and $16$ \cite[Ch.~1]{conway99}. For this
reason, they have been applied in a number of applications, including digital
communications \cite{forney88pt1}, data compression \cite{adoul95},
cryptography \cite{lyu22pke},
quantum computing \cite{calderbank1998quantum},
and
algebraic geometry \cite{buser1994period}.

The Voronoi regions of $D_4$ and $E_8$ have been fully determined. Hence their
\emph{packing densities,} \emph{kissing numbers,} and \emph{quantizer constants}
are known exactly
\cite{conway82}, \cite[Ch.~4]{conway99}, and we will not discuss
these lattices further. In this paper, we determine the Voronoi region of the
$16$-dimensional BW lattice $\BW$.
Its relevant vectors, vertices and quantizer constant are
reported exactly for the first time.
We furthermore characterize
its full symmetry group,
which is known to be of order
$89\,181\,388\,800$ \cite[Section~4.10]{conway99},
using two transformation matrices.

\section{The face hierarchy}

In this section, we describe the Voronoi region of $\BW$ in a
bottom-up manner, beginning from the $0$-faces (vertices) and making our way
upwards in the hierarchy of dimensions to the single $16$-face, which is the
Voronoi region itself. We describe the faces in the coordinate system defined
by the lower block triangular generator matrix
\begin{align} \label{eq:gen}
\newcommand{\h}{\frac{1}{2}}
\renewcommand\arraystretch{\IfElseFinal{1.1}{0.7}}
\left[\begin{array}{@{\!\!}*{8}{@{\ }c@{\ }};{2pt/2pt}@{}c@{}*{8}{@{\ }c@{\ }}@{\!\!}}
2  & 0  & 0  & 0  & 0  & 0  & 0  & 0  && 0  & 0  & 0  & 0  & 0  & 0  & 0  & 0 \\
1  & 1  & 0  & 0  & 0  & 0  & 0  & 0  && 0  & 0  & 0  & 0  & 0  & 0  & 0  & 0 \\
1  & 0  & 1  & 0  & 0  & 0  & 0  & 0  && 0  & 0  & 0  & 0  & 0  & 0  & 0  & 0 \\
1  & 1  & 1  & 1  & 0  & 0  & 0  & 0  && 0  & 0  & 0  & 0  & 0  & 0  & 0  & 0 \\
1  & 0  & 0  & 0  & 1  & 0  & 0  & 0  && 0  & 0  & 0  & 0  & 0  & 0  & 0  & 0 \\
1  & 1  & 0  & 0  & 1  & 1  & 0  & 0  && 0  & 0  & 0  & 0  & 0  & 0  & 0  & 0 \\
1  & 0  & 1  & 0  & 1  & 0  & 1  & 0  && 0  & 0  & 0  & 0  & 0  & 0  & 0  & 0 \\
\rule[-1.5ex]{0pt}{0pt}
\h & \h & \h & \h & \h & \h & \h & \h && 0  & 0  & 0  & 0  & 0  & 0  & 0  & 0 \\
\hdashline[2pt/2pt]
\tstrut
1  & 0  & 0  & 0  & 0  & 0  & 0  & 0  && 1  & 0  & 0  & 0  & 0  & 0  & 0  & 0 \\
1  & 1  & 0  & 0  & 0  & 0  & 0  & 0  && 1  & 1  & 0  & 0  & 0  & 0  & 0  & 0 \\
1  & 0  & 1  & 0  & 0  & 0  & 0  & 0  && 1  & 0  & 1  & 0  & 0  & 0  & 0  & 0 \\
\h & \h & \h & \h & 0  & 0  & 0  & 0  && \h & \h & \h & \h & 0  & 0  & 0  & 0 \\
1  & 0  & 0  & 0  & 1  & 0  & 0  & 0  && 1  & 0  & 0  & 0  & 1  & 0  & 0  & 0 \\
\h & \h & 0  & 0  & \h & \h & 0  & 0  && \h & \h & 0  & 0  & \h & \h & 0  & 0 \\
\h & 0  & \h & 0  & \h & 0  & \h & 0  && \h & 0  & \h & 0  & \h & 0  & \h & 0 \\
\h & \h & \h & \h & \h & \h & \h & \h && \h & \h & \h & \h & \h & \h & \h & \h
\end{array}\right]
\;.
\end{align}
This generator matrix is scaled down by a linear factor of $\surd 2$
(or, equivalently, a volume factor of 256)
compared with the generator matrix for the same lattice in \cite[Fig.~4.10]{conway99}.
Some lattice parameters depend on the scaling of the lattice.

\subsection{0-faces}
The Voronoi region has
$201\,343\,200$ vertices,
which belong to six equivalence classes listed
as $\vc v_1,\vc v_2,\ldots,\vc v_6$ in Tab.~\ref{tab:BW16-normals-vertices}.
Equivalence is defined by the rotations $\grp$
that take $\BW$ into $\BW$.
If translation by a lattice vector is considered as another equivalence
operation,
$\vc v_2$ becomes equivalent to $\vc v_4$ and $\vc v_3$ to $\vc v_5$,
reducing the six equivalence classes to only four.
The vertices are located at a squared distance from the origin of $3/2$,
$10/9$, or $1$.
Hence, the covering radius is $\sqrt{3/2}$, as already known
\cite{forney88pt2}, \cite[Section~4.10]{conway99}.

\begin{centeredtable}{%
    Representatives of the relevant vectors $\vc n_i$ (first two rows)
    and vertices $\vc v_i$ (remaining six rows)
    of the Voronoi region of $\BW$ in order of increasing length.
    Shown are the components, squared lengths, sizes of the orbits under $\grp$,
    and sizes of the respective stabilizer subgroups of $\grp$.
}
    \label{tab:BW16-normals-vertices}
    \begin{tabular}{l@{\quad}r*{14}{@{\IfElseFinal{ }{\;}}r@{\IfElseFinal{ }{\;}}}r@{\quad}c@{\quad}rr} \hline
        \rule{0pt}{2.1ex}%
        vector & \multicolumn{16}{c}{components} & $\| \cdot\|^2$ & orbit size & stabilizer size \\
        \hline \tstrut
        $\vc n_{1}$ & $(1$                & $ \phantom{-}1$ & $ 0$ & $ 0$ & $ 0$ & $ 0$ & $ 0$ & $ 0$ & $\phantom{-}0$ & $ 0$ & $ 0$ & $ 0$ & $ 0$ & $ 0$ & $ 0$ & $ 0)$ & $2$ & $4\,320$  & $20\,643\,840$ \\
        \bstrut
        $\vc n_{2}$ & ${1}/{2}$ $( 2$ & $ 1$            & $ 1$ & $ 0$ & $ 1$ & $ 0$ & $ 0$ & $ 1$ & $ 1$           & $ 0$ & $ 0$ & $ 1$ & $ 0$ & $ 1$ & $-1$ & $ 0)$ & $3$ & $61\,440$ & $1\,451\,520$ \\
        \hline \tstrut
        $\vc v_{1}$ & ${1}/{2}$ $( 1$  & $ 1$ & $ 0$ & $ 0$ & $ 0$ & $ 0$ & $ 0$ & $ 0$ & $ 0$ & $ 0$ & $ 0$ & $ 0$ & $ 0$ & $ 0$ & $ 1$ & $ 1)$ & $1$            & $4\,320$       & $20\,643\,840$ \\
        $\vc v_{2}$ & ${1}/{12}$ $( 9$ & $ 3$ & $ 3$ & $ 3$ & $ 1$ & $-1$ & $ 1$ & $ 1$ & $ 1$ & $-1$ & $ 3$ & $-1$ & $-3$ & $ 3$ & $-3$ & $-3)$ & ${10}/{9}$ & $66\,355\,200$ & $1\,344$ \\
        $\vc v_{3}$ & ${1}/{6}$ $( 5$  & $ 1$ & $-1$ & $-1$ & $-1$ & $ 1$ & $ 1$ & $-1$ & $ 1$ & $-1$ & $ 1$ & $-1$ & $ 1$ & $ 1$ & $ 1$ & $ 1)$ & ${10}/{9}$ & $2\,211\,840$  & $40\,320$ \\
        $\vc v_{4}$ & ${1}/{6}$ $( 5$  & $ 1$ & $ 1$ & $ 1$ & $-1$ & $ 1$ & $-1$ & $-1$ & $ 1$ & $-1$ & $-1$ & $-1$ & $-1$ & $-1$ & $ 1$ & $-1)$ & ${10}/{9}$ & $66\,355\,200$ & $1\,344$ \\
        $\vc v_{5}$ & ${1}/{6}$ $( 5$  & $ 1$ & $ 1$ & $ 1$ & $ 1$ & $ 1$ & $-1$ & $-1$ & $ 1$ & $ 1$ & $ 1$ & $-1$ & $ 1$ & $ 1$ & $-1$ & $ 1)$ & ${10}/{9}$ & $66\,355\,200$ & $1\,344$ \\
        \bstrut
        $\vc v_{6}$ & ${1}/{4}$ $( 3$  & $ 1$ & $ 1$ & $ 1$ & $ 1$ & $ 1$ & $ 1$ & $-1$ & $ 1$ & $ 1$ & $ 1$ & $-1$ & $ 1$ & $-1$ & $-1$ & $-1)$ & ${3}/{2}$  & $61\,440$      & $1\,451\,520$ \\
        \hline
    \end{tabular}
\end{centeredtable}

\subsection{1-faces}
The vertices are connected by a total of
about $3\cdot10^{10}$
edges, which belong to $23$
equivalence classes.\footnote{
    We explicitly construct at least one representative per equivalence class.
    However, we do not create the full orbits of those representatives.
    The number of faces is instead obtained by estimating the
    orbit sizes using \cite{orb4-8-3}.
}
Their lengths are
$\sqrt{3/2}$,
$1$,
$\sqrt{17/18}$,
$\sqrt{7}/3$,
$\sqrt{11/18}$,
$2/3$,
$\sqrt{5/18}$,
and
$1/3$.
At each vertex equivalent to
$\vc v_1$,
$\vc v_2$,
$\vc v_3$,
$\vc v_4$,
$\vc v_5$, or
$\vc v_6$,
respectively,
$32\,768$,
$144$,
$403$,
$179$,
$220$,
or
$398\,824$
edges meet.

\subsection{2-faces}
\begin{figure*}
    \begin{center}
        \includegraphics[width=1\linewidth]{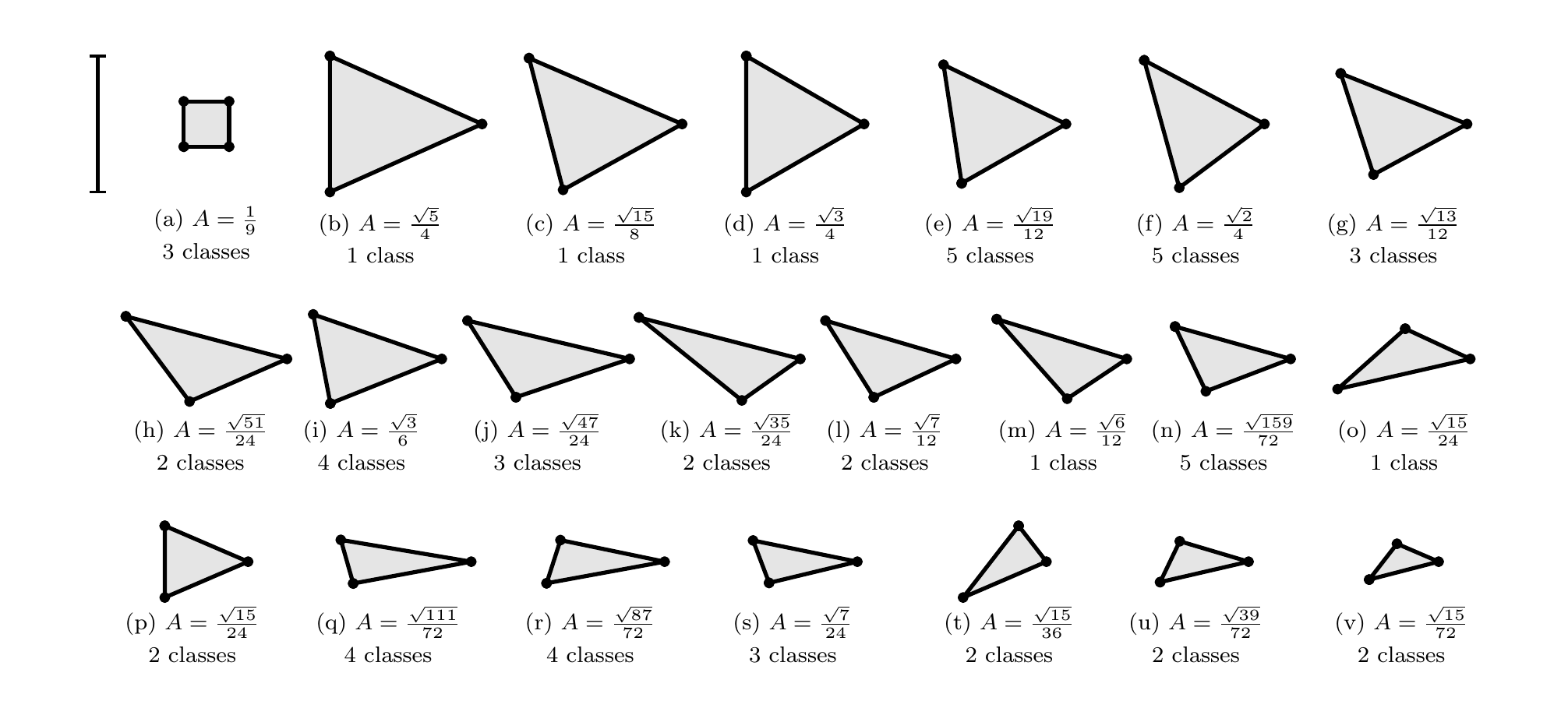}
    \end{center}
    \caption{\label{fig:2-faces}
        The $22$ geometrically distinct types of $2$-faces of the
        Voronoi region of $\BW$.
        We show the area $A$ as well as the number of classes the face types fall
        into under $\grp$.
        For reference, the line in the top-left corner has unit length.
        (a)~is a square with edges of length $1/3$.
        The triangle~(d) is equilateral, while
        (b),
        (c),
        (e),
        (g),
        (h),
        (l),
        (p),
        (t),
        and
        (v)
        are isosceles.
        All internal angles of the triangles are strictly less than $90^\circ$.
        The maximum $\theta_\text{max}$ is only found in (n), where
        $\cos\theta_\text{max} = \sqrt{10}/40$
        ($\theta_\text{max} \approx 85.47^\circ$).
        The smallest angle satisfies
        $\cos\theta_\text{min} = 29\sqrt{238}/476$
        ($\theta_\text{min} \approx 19.97^\circ$)
        and is only found in triangle (q).
    }
\end{figure*}

There are
about $5\cdot10^{11}$
$2$-faces in $58$ equivalence classes.
These consist of
$3$ classes of
about $4\cdot10^{9}$
squares with an area of $1/9$
and about $5\cdot10^{11}$ triangles in $55$ classes,
$21$ of which are geometrically distinct,
with areas between $\sqrt{15}/72$ and $\sqrt{5}/4$.
The $22$ geometrically distinct $2$-faces are shown in
Fig.~\ref{fig:2-faces}.

\subsection{3- to 14-faces}
In dimensions $3$ to $14$, there are
$6\,052$ classes of faces, which we will not describe in detail here.
Some of their properties are summarized in Tab.~\ref{tbl:hist},
where we show the number of face classes under $\grp$,
numbers of child faces (i.e., subfaces of dimension $d-1$) and vertices for the
faces in all dimensions $d = 0,1,\ldots,16$.
Further information is available as supplementary material
\cite{ancillary-files}.

\begin{centeredtable}{
        Summary information about the faces of the Voronoi region of $\BW$.
        The first column lists the dimension $d$ of the faces and
        the second the number of classes of $d$-faces under $\grp$.
        The third column shows the range of numbers of child faces
        of each $d$-face
        and the fourth column the range of numbers of vertices of each
        $d$-face.
        In the fifth column, we visualize the number of
        face classes ($y$-axis) containing a certain number of vertices
        ($x$-axis).
        The last column shows the same information for the numbers of faces
        instead of face classes, which have been approximated using
        \cite{orb4-8-3}.
    }
    \label{tbl:hist}
    \IfElseFinal{%
        \def \incplot #1{\raisebox{0.75em}{\includegraphics[scale=0.75,valign=t]{#1}}}
    }{%
        \def \incplot #1{\raisebox{0.75em}{\includegraphics[scale=0.65,valign=t]{#1}}}
        \scriptsize
    }%
    \begin{tabular}{r|ccccc}
        \hline
            \tstrut
            dim & classes & child faces & vertices
            & \hspace{\IfElseFinal{2em}{1.7em}} vertex counts of face classes
            & \hspace{\IfElseFinal{2.8em}{2.0em}} vertex counts of faces (approx.) \\
        \hline
            \rule{0pt}{4.4ex}
            $0$  & $6$      & $0$             & $1$                      & \incplot{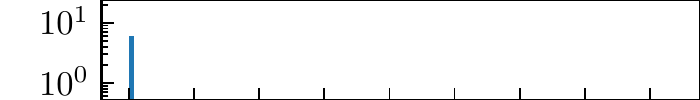}  & \incplot{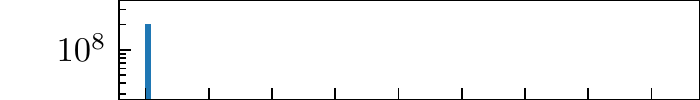} \\
            $1$  & $23$     & $2$             & $2$                      & \incplot{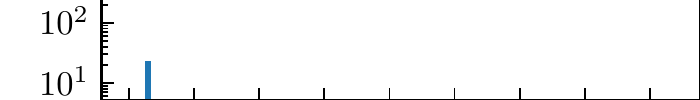}  & \incplot{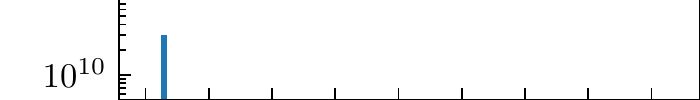} \\
            $2$  & $58$     & $3$, $4$        & $3$, $4$                 & \incplot{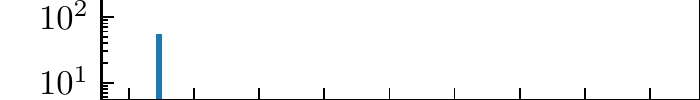}  & \incplot{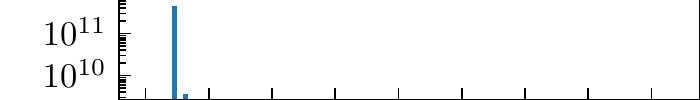} \\
            $3$  & $168$    & $4$--$6$        & $4$--$8$                 & \incplot{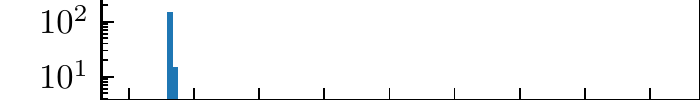}  & \incplot{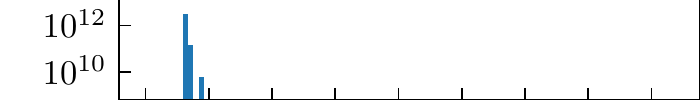} \\
            $4$  & $441$    & $5$--$16$       & $5$--$16$                & \incplot{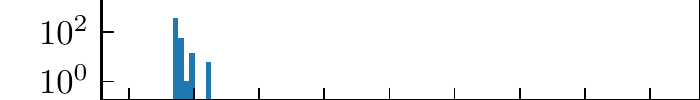}  & \incplot{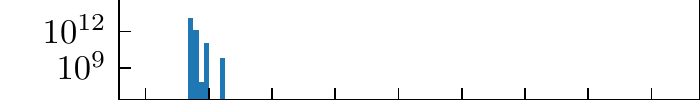} \\
            $5$  & $867$    & $6$--$21$       & $6$--$32$                & \incplot{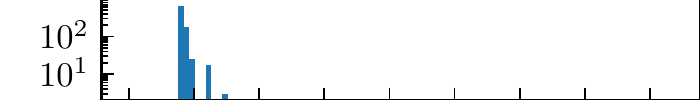}  & \incplot{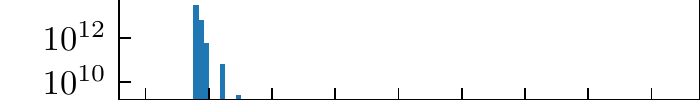} \\
            $6$  & $1\,257$ & $7$--$30$       & $7$--$64$                & \incplot{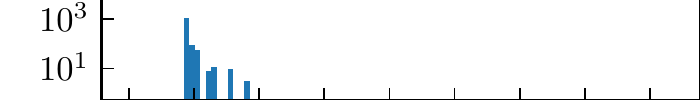}  & \incplot{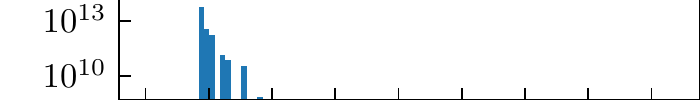} \\
            $7$  & $1\,329$ & $8$--$51$       & $8$--$128$               & \incplot{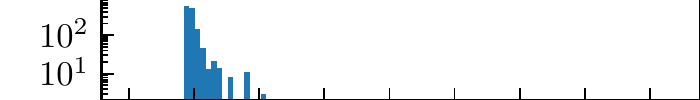}  & \incplot{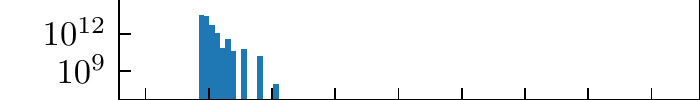} \\
            $8$  & $1\,023$ & $9$--$128$      & $9$--$256$               & \incplot{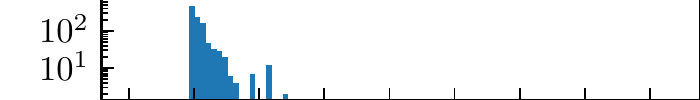}  & \incplot{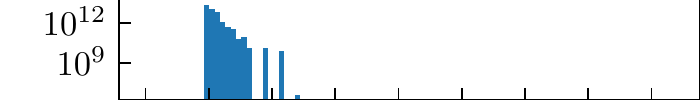} \\
            $9$  & $566$    & $10$--$194$     & $10$--$400$              & \incplot{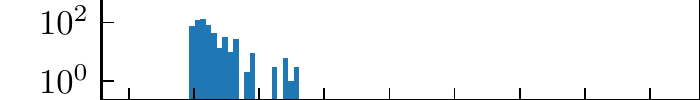}  & \incplot{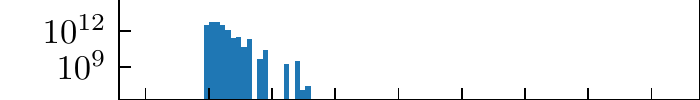} \\
            $10$ & $253$    & $11$--$258$     & $11$--$641$              & \incplot{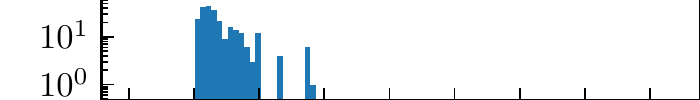} & \incplot{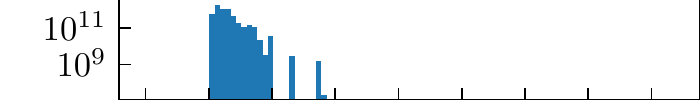} \\
            $11$ & $96$     & $12$--$620$     & $12$--$1\,281$           & \incplot{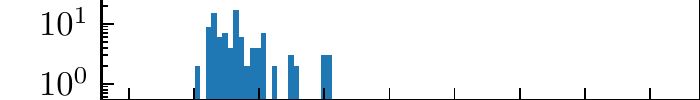} & \incplot{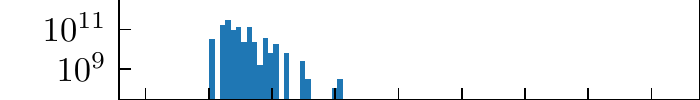} \\
            $12$ & $35$     & $16$--$862$     & $24$--$2\,945$           & \incplot{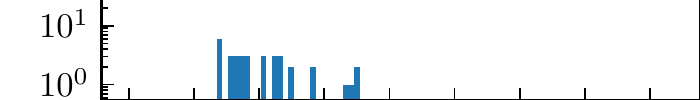} & \incplot{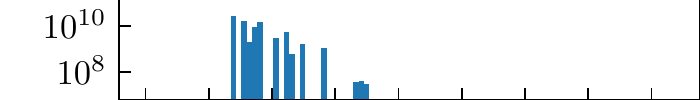} \\
            $13$ & $12$     & $42$--$1\,312$  & $64$--$11\,138$          & \incplot{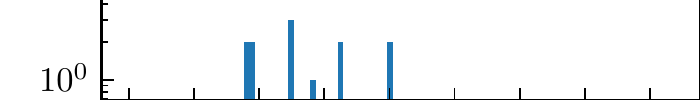} & \incplot{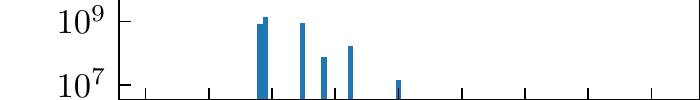} \\
            $14$ & $5$      & $144$--$2\,763$ & $520$--$59\,907$         & \incplot{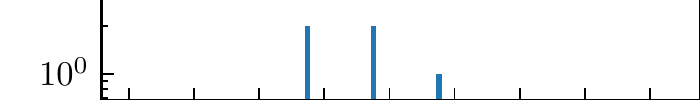} & \incplot{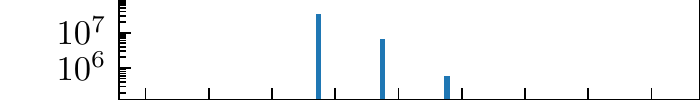} \\
            $15$ & $2$      & $828$, $7\,704$ & $26\,160$, $1\,046\,430$ & \incplot{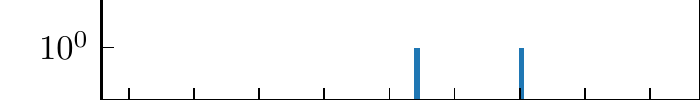} & \incplot{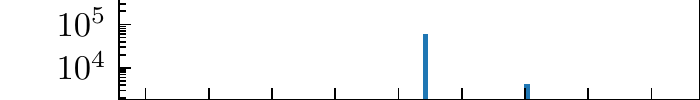} \\
            \rule[\IfElseFinal{-8.5ex}{-6.9ex}]{0pt}{0pt}
            $16$ & $1$      & $65\,760$       & $201\,343\,200$          & \incplot{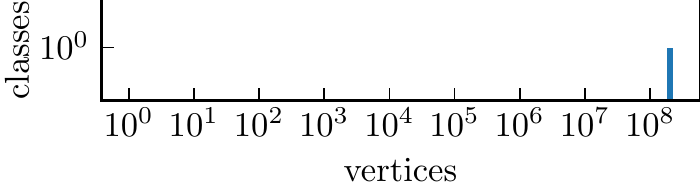} & \incplot{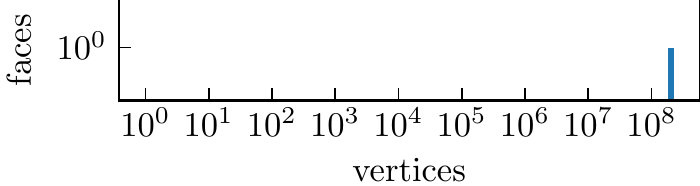} \\
        \hline
    \end{tabular}
\end{centeredtable}

\subsection{15-faces}
The $15$-faces, or \emph{facets,}
all lie halfway between the origin and another lattice vector,
orthogonal to the line between them.
There are in total $65\,760$ such facet-defining nonzero vectors, or
\emph{relevant vectors.}
They belong to two equivalence classes at
different distances from the origin (see Tab.~\ref{tab:BW16-normals-vertices}).
The ones closest to the origin are the \emph{minimal vectors} at a
squared distance of $2$,
which were found already in \cite{barnes59}.
The packing radius is half of their length, i.e., $\sqrt{2}/2$.
There are $4\,320$ such vectors, which is the kissing number of the lattice.
There are also $61\,440$ other relevant vectors, which have a squared length of $3$.

The facets belonging to the $4\,320$ minimal vectors each have
$7\,704$ child faces and
$1\,046\,430$ vertices of all six classes,
while the remaining $61\,440$ facets have
$828$ child faces and
$26\,160$ vertices equivalent
to either $\vc v_2$, $\vc v_4$, $\vc v_5$, or $\vc v_6$.

\subsection{16-face}

Having enumerated all inequivalent $d$-faces for $d=0,1,\ldots,15$ and
computed their volumes and second moments using the recursion relations in
\cite[Sec.~3]{Pook-Kolb2022Exact}, a complete characterization of the $16$-face
is obtained. Using \cite{orb4-8-3}, we estimate
that the Voronoi region has between
$1\cdot10^{14}$ and $3\cdot10^{14}$ faces across all dimensions.

Next, the {\em covariance matrix} or {\em second moment tensor} is computed as
\begin{equation}\label{eq:BW16-Uab}
    \mat U = \frac{U}{16} \mat I_{16}
        \;,
\end{equation}
where the (unnormalized) \emph{second moment}
\begin{equation}\label{eq:BW16-U}
    U = \tr \mat U = \frac{207\,049\,815\,983}{4\,287\,303\,820\,800}
\end{equation}
and $\mat I_{16}$ the $16 \times 16$ identity matrix.
After proper normalization, the quantizer constant is obtained as
\begin{equation}\label{eq:G-formula}
    G \defeq \frac{1}{n} \frac{U}{V^{1+2/n}}
    \;,
\end{equation}
where $n = 16$ is the lattice's dimension and $V = 1/16$ is the volume of
its Voronoi region,
which yields
\begin{equation}\label{eq:BW16-G}
    G = U\sqrt{2}
    \approx 0.068\,297\,622\,489\,318\,7
    \;.
\end{equation}

To verify our enumeration of face classes, we use the
recursion relations in \cite[Sec.~3]{Pook-Kolb2022Exact} to calculate the
volume of the Voronoi region, which agrees with the expected value of $1/16$.
We also verify the result \eqnref{eq:BW16-G} numerically
in Sec.~\ref{sec:numerics}.

\section{The symmetry group of \texorpdfstring{$\BW$}{Lambda16}}
\label{sec:symmetries}

The symmetries of $\BW$ are generated by products of sign changes, permutations and the matrix
\begin{equation}\label{eq:Hsym}
    \mat H \defeq \begin{bmatrix}
        \mat H_4 & \mat 0   & \mat 0   & \mat 0\\
        \mat 0   & \mat H_4 & \mat 0   & \mat 0\\
        \mat 0   & \mat 0   & \mat H_4 & \mat 0\\
        \mat 0   & \mat 0   & \mat 0   & \mat H_4
    \end{bmatrix}
    \;,
\end{equation}
where
\begin{equation}\label{eq:H4}
    \mat H_4 \defeq \frac{1}{2} \left[\begin{array}{@{\,}rrrr@{\,}}
        1 & 1 & 1 & 1\\
        1 &-1 & 1 &-1\\
        1 & 1 &-1 &-1\\
        1 &-1 &-1 & 1
    \end{array}\right]
\end{equation}
is a Hadamard matrix.

There are $2\,048$ sign changes, which can be described as a product of three subgroups
$\grpS_1$, $\grpS_2$ and $\grpS_3$.
The first subgroup $\grpS_1$ contains all even numbers of sign changes of
component pairs $(\vc x_i, \vc x_{i+1})$ for $i = 1, 3, \ldots 15$,
and has order $128$.
$\grpS_2$ changes the signs of an even number of the first and last $4$ odd components
$(\vc x_i, \vc x_{16-i})$, $i = 1, 3, 5, 7$.
This subgroup has order $8$.
Finally, $\grpS_3$ is of order $2$ and changes the signs of
the components $(\vc x_1, \vc x_3, \vc x_5, \vc x_7)$.

The permutations $\grpP \subset \grp$ of vector components that keep $\BW$ invariant
are described in \cite[Lemma~3.2]{barnes59}.
The Lemma makes use of a $4$-dimensional vector space over the
Galois field $\mathrm{GF}(2)$ to represent indices of components of the
lattice vectors.
The reader is referred to \cite{barnes59} for a detailed description of this
construction.
Using
\cite[Eq.~(19) of Ch.~13]{macwilliams1977theory}
the order of $\grpP$ is
\begin{equation}\label{eq:P-order}
    |\grpP| = 16 \prod_{l=0}^3 (16 - 2^l) = 322\,560
    \;,
\end{equation}
These are precisely the permutations that keep the
first-order binary Reed--Muller codes of length $2^4$ invariant
\cite[Theorem~24 of Ch.~13]{macwilliams1977theory}.

Examples of permutations in $\grpP$ are
\begin{align}
    \notag
    p_1 &= ( 1\ 2\ 3\ 4)( 5\ 6\ 7\ 8)( 9\ 10\ 11\ 12)(13\ 14\ 15\ 16) \;, \\
    \notag
    p_2 &= ( 1\ 2)( 5\ 6)( 9\ 10)(13\ 14) \;, \\
    \notag
    p_3 &= ( 1\  6\ 13)( 2\  8)( 3\  9\ 12\  5\ 15\ 14)( 4\ 11\  7) \;, \\
    \label{eq:example-perms}
    p_4 &= (1\ 9\ 16\ 15\ 5\ 7\ 4\ 8\ 10\ 6\ 13\ 2\ 3\ 14\ 12)
    \;,
\end{align}
here given in cycle notation for compactness.
The complete subgroup $\grpP$ can be generated using various subsets of these
permutations, for example
$\{p_1, p_2, p_3\}$,
$\{p_1, p_4\}$,
or
$\{p_3, p_4\}$.

The full automorphism group $\grp$
can be generated by combining
$\mat H$
with the generators of $\grpS_1$, $\grpS_2$, and $\grpS_3$
and one of the sets of generators of $\grpP$.
Remarkably, it can also be generated by just two matrices.
The first is the $16 \times 16$ permutation matrix $\mat M_1$ corresponding to $p_3$.
The second is a matrix
\begin{equation}\label{eq:M2}
    \mat M_2 \defeq \begin{bmatrix}
        \bar{\mat H}_4 & \mat 0        & \mat 0        & \mat 0\\
        \mat 0        & \bar{\mat H}_4 & \mat 0        & \mat 0\\
        \mat 0        & \mat 0        & \bar{\mat H}_4 & \mat 0\\
        \mat 0        & \mat 0        & \mat 0        & \bar{\mat H}_4
    \end{bmatrix}
    \;,
\end{equation}
which is built using \eqnref{eq:H4} with a sign change of the last row, i.e.,
with the Hadamard matrix
\begin{equation}\label{eq:H4bar}
    \bar{\mat H}_4 \defeq \frac{1}{2} \left[\begin{array}{@{\,}rrrr@{\,}}
        1 & 1 & 1 & 1\\
        1 &-1 & 1 &-1\\
        1 & 1 &-1 &-1\\
       -1 & 1 & 1 &-1
    \end{array}\right] \;.
\end{equation}

\section{Numerical verification and error estimates}
\label{sec:numerics}

To validate \eqref{eq:BW16-U}, we estimate $U$ by Monte-Carlo integration over
the Voronoi region. We also estimate the variance of
the estimate of $U$, for which we use a different method than the
``jackknife estimator'' in \cite{conway84}.
In this section, we first describe our
estimate of $U$ and the variance thereof, then motivate why we prefer
our variance estimator over the jackknife, and finally compare our numerical
estimate of $G$ for $\BW$ with the true value in \eqref{eq:BW16-G}.

The Monte-Carlo estimate of $U$ is
\begin{equation}\label{eq:uhat}
\hat{U} = \frac{1}{N}\sum_{i=1}^N \|\vc x_i \|^2
\;,
\end{equation}
where $\vc x_1,\ldots, \vc x_N$ are $N$ independent random vectors
uniformly distributed in the Voronoi region of $\Lambda$.

To estimate $\var\hat{U}$, we first note that since the vectors $\vc x_i$ are
independent and identically distributed,
$\var\hat{U} = (1/N) \var\!\|\vc x\|^2$, where $\vc x$ is a single random vector
with the same distribution as $\vc x_i$.
Therefore, our estimate
of $\var\hat{U}$, denoted by $\varh\hat{U}$,
is defined by
\begin{equation}\label{eq:varhat}
    \varh\hat{U} = (1/N) \varh\!\|\vc x\|^2
    \;.
\end{equation}
Applying
the standard unbiased variance estimator
of $\var\!\|\vc x\|^2$
\begin{equation}
\varh\!\|\vc x\|^2 = \frac{1}{N-1}\sum_{i=1}^N \left(\|\vc x_i \|^2-\hat{U}\right)^2
\end{equation}
in \eqref{eq:varhat} yields
\begin{align} \label{eq:var-est}
\varh\hat{U} &= \frac{1}{N(N-1)}\sum_{i=1}^N \left(\|\vc x_i \|^2-\hat{U}\right)^2 \notag\\
  &= \frac{1}{N-1} \left(\frac{1}{N}\sum_{i=1}^N\|\vc x_i \|^4-\hat{U}^2\right)
\end{align}
or after normalization as in \eqref{eq:G-formula}
\begin{align}
    \hat{G} &= \frac{\hat{U}}{nV^{1+2/n}}, \label{eq:ghat} \\
    \varh{\hat{G}} &= \frac{\varh{\hat{U}}}{(nV^{1+2/n})^2} \label{eq:G-var}
\;.
\end{align}

\begin{figure}
\centering
\includegraphics[width=\IfElseFinal{1.0}{0.6}\columnwidth]{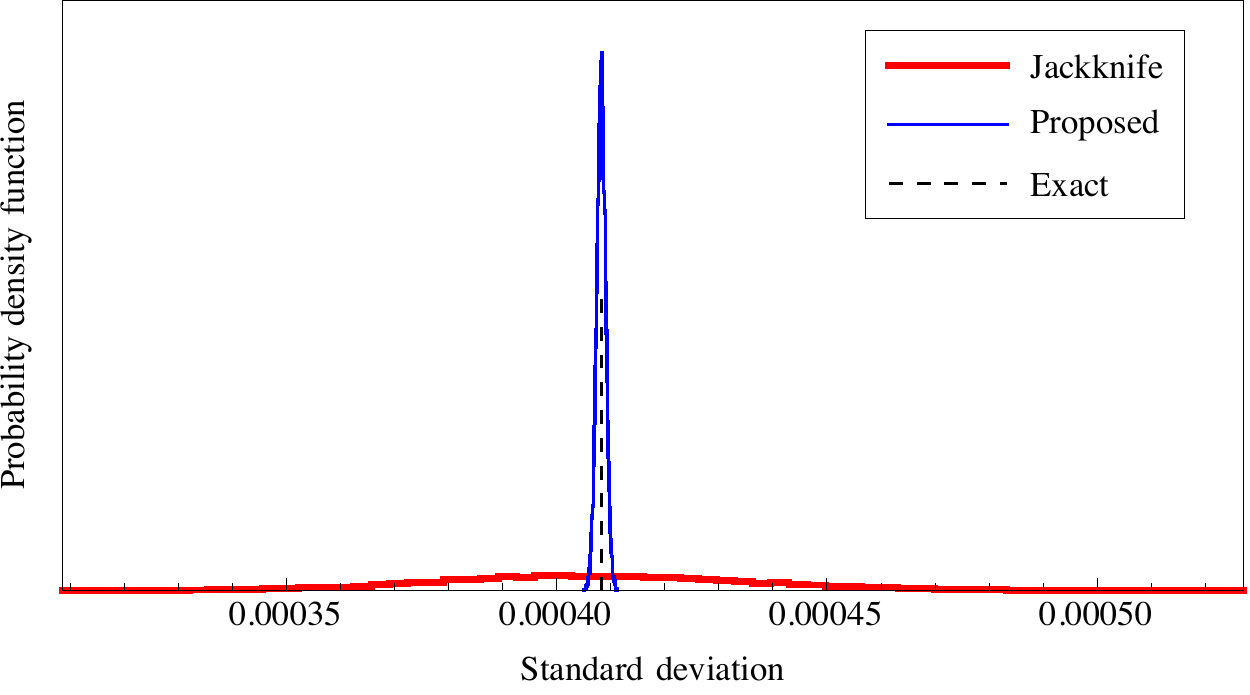}
\caption{Histograms of two estimates of the standard deviation of the
estimated second moment $\hat{U}$ of the cubic lattice.
The exact standard deviation $(\var \hat{U})^{1/2}$, which can be calculated analytically
for the cubic lattice,
reveals that the proposed estimator \eqref{eq:uhat} is much more accurate than the jackknife
with 100 groups.}
\label{fig:U-pdfs}
\end{figure}

The variance estimator \eqref{eq:var-est} follows directly from fundamental
laws of probability.
What is surprising is that a different estimator has been used, unchallenged, in most, or
perhaps all, previous works involving numerical estimates of lattice second moments
\cite{conway84, agrell98, Lyu2022better}.
To rectify this 39-year old misconception, we now elaborate on why
\eqref{eq:var-est} is more accurate.

The jackknife works by partitioning the independent
randomly selected
vectors $\vc x_1,\ldots, \vc x_N$
into $g$ groups, computing the average squared length within each group,
and finally computing the sample variance of these $g$ averages \cite[Eqs.~(3)--(4)]{conway84}.
This method brings at least two disadvantages:
First, the estimated variance depends on how the list $\vc x_1,\ldots, \vc x_N$ is ordered;
reordering the list would yield a different variance estimate,
although the estimated second moment \eqref{eq:uhat} remains the same.
And second, the variance of vectors within a group is ignored.
The proposed estimator \eqref{eq:var-est} suffers from neither of these disadvantages.

To quantify the accuracy of both variance estimators,
we numerically estimate the second moment of the cubic lattice $\Z^n$ for $n=3$.
The second moment of $\Z^n$ is $U = \E[\|\vc x\|^2] = n/12$,
and the variance of $\hat{U}$ can be calculated exactly as
$\var \hat{U} = (1/N) \var\!\|\vc x\|^2 = (1/N)(\E[\|\vc x\|^4] - \E[\|\vc x\|^2]^2) = n/(180N)$.
We generated $N=100\,000$ vectors uniformly in the Voronoi region of $\Z^3$,
which is the unit cube,
computed $\hat{U}$ using \eqref{eq:uhat},
and estimated the variance of $\hat{U}$ using the two methods.
For the jackknife, we used a group size of $g=100$ as in \cite{conway84}.
Both estimators were run 10\,000 times, each time with $N$ new random vectors.
Fig.~\ref{fig:U-pdfs} shows histograms of the resulting estimates of the
standard deviation, together with the exact value.
It can be observed that \eqref{eq:uhat} in this example is more than an order
of magnitude more accurate than the jackknife with $g=100$.

The accuracy of the jackknife improves with increasing $g$, and it is most
accurate when each group consists of a single sample, i.e., when $g=N$.
In this extreme case, the jackknife simplifies into \eqref{eq:var-est}---but
this is not how the jackknife was applied in previous studies
\cite{conway84, agrell98, Lyu2022better}.

Having established the usefulness of the new variance estimator,
we proceed to estimate the quantizer constant $G$ of $\BW$ with high accuracy.
Numerically evaluating \eqref{eq:uhat} and \eqref{eq:ghat} for the
mean and \eqref{eq:var-est} and \eqref{eq:G-var} for the standard deviation,
using $N = 4\cdot 10^{12}$ random $16$-dimensional
vectors, we obtain
\begin{align}
\hat{G} &= 0.068297616
\;, \\
\sqrt{\varh \hat{G}} &= 0.000000009
\;.
\end{align}
The difference between $\hat{G}$ and the exact $G$ in \eqref{eq:BW16-G} is only $0.7$
standard deviations, which may serve as a numerical verification of the face hierarchy.
The results are also in agreement with the previous (less accurate) estimate of the
same constant in \cite[Eq.~(13)]{conway84}.

\section{The algorithm}
\label{sec:alg}

Our algorithm\footnote{%
    The algorithms are implemented in \lib{Python} and the data types \class{List} and
    \class{Dictionary} we use in the code listings are meant to behave like the
    respective Python types.
    Group-theoretic aspects make use of \lib{GAP}
    \cite{GAPSoftware,GAPRepo}, which is called from Python using
    \lib{gappy} \cite{gappySoftware}.
}
is described in detail in \cite{Pook-Kolb2022Exact},
which builds on previous
methods for finding all relevant vectors \cite{AEVZ} and faces \cite{allen21}.
In this section, we briefly summarize the main concept and present
minor modifications to
the methods of \cite{Pook-Kolb2022Exact}.

The basic approach remains the same:
We first find all relevant vectors, i.e., normals of the facets,
and all the vertices of the Voronoi region.
The hierarchy of subfaces of the facets is then built by recursively
intersecting the sets of vertices of parent faces.
The computational cost is kept low by finding the classes of faces
equivalent under $\Aut(\BW)$ and then only constructing the child faces of one
(arbitrarily chosen) representative face per class.
In total, only $159\,143$ faces are constructed explicitly.

The classification of faces is performed iteratively
as described in \cite[Section~2.4.4]{Pook-Kolb2022Exact}.
In this method, we begin identifying equivalent faces using a proper subgroup
$\setU \subset \grp$,
which creates classes of faces under $\setU$.
The set consisting of one (arbitrary) representative per class
is then classified using another subgroup $\setU'$.
This can be repeated with different subgroups until we finally use
the full group $\grp$.
For $\BW$, we found that a good option is to use only a
single subgroup $\setU$, chosen as the stabilizer
of the relevant vector $\vc n_2$ with a stabilizer size of $1\,451\,520$
(see Tab.~\ref{tab:BW16-normals-vertices}).

We made three changes to the method in
\cite{Pook-Kolb2022Exact},
which affect
how the equivalence of two faces is tested
and
how the orbits and stabilizers of individual vectors are constructed.
We now describe these changes in turn, briefly revisiting the respective previous
methods followed by our new algorithms.

\subsection{Testing the equivalence of faces}
\label{sub:face-eq}

Our previous method of testing whether
a face $F$ is equivalent to another face $F'$ under a group $\grpG$
is based on the following idea.\footnote{
    Here, $\grpG$ is either $\grp$ or $\setU$.
}
For each face, we take a set of vectors that uniquely identifies that
face.
We use either the set of relevant vectors associated with the facets
containing the face (i.e., the ``normal vectors'' of the face)
or alternatively the face's vertices.
The choice depends on the number of vectors in either of the two sets and
on their classification under $\grpG$.
Let $\vc x_1, \ldots, \vc x_N$ be the vectors of $F$ and
$\vc y_1, \ldots, \vc y_N$ be those of $F'$.
We order these vectors such that $\vc x_i$ is equivalent to $\vc y_i$ for all $i$
(if that is not possible, the faces are inequivalent).
We then form the sets of all transformations
between pairs $(\vc x_i, \vc y_i)$ for all $i$.
If the intersection of these sets is non-empty, it consists of transformations
taking $F$ into $F'$.
If it is empty, however, we permute one of the sets and try again.
The faces are inequivalent if and only if all permutations lead to empty
intersections of the sets of transformations.

In principle, the full set of transformations between any two equivalent
vectors
can easily be constructed as follows.
Let $\vc x = g_x \xrep$ and $\vc y = g_y \xrep$ be two equivalent vectors with
$g_x, g_y \in \grpG$ and $\xrep$ representing their equivalence class.
Then, the full set of transformations in $\grpG$ taking $\vc x$ into $\vc y$
is \cite{Pook-Kolb2022Exact}
\begin{equation}\label{eq:Txy}
    \setT_{xy} =
    g_y \Stab_\grpG(\xrep) g_x^{-1}
    \;,
\end{equation}
where $\Stab_\grpG(\xrep)$ is the stabilizer of $\xrep$ in $\grpG$.

From Tab.~\ref{tab:BW16-normals-vertices}, we see that
for $\BW$, the sets \eqnref{eq:Txy}
contain between $1\,344$ and $20\,643\,840$ elements.
When forming the intersections using \lib{GAP},
these sets are held in memory, which becomes a problem when multiple
intersections need to be calculated.

\begin{alg}
    \caption{
        Evaluate if two faces $F$ and $F'$ are equivalent under $\grpG$.
        If they are, return a transformation $g \in \grpG$ taking $F$ into $F'$,
        otherwise return NULL.
        We define $\lvert\XX\rvert$
        as the number of elements in a set $\XX$ and
        $\argmin_{\vc x \in \XX} f$ as the function returning the subset of $\XX$
        for which $f(\vc x)$ is smallest.
    }\label{alg:FindTransformation}
    \begin{algorithmic}[1]
        \Procedure{FindTransformation}{$F$, $F'$, $\grpG$}
            \State $\verts$ $\gets$ vertices of $F$
            \State $\verts'$ $\gets$ vertices of $F'$
            \State $\normals$ $\gets$ normal vectors of $F$
            \State $\normals'$ $\gets$ normal vectors of $F'$
            \CommentLineEmpty[1]{There may be multiple minima, which are all captured:}
            \State $\setM$ $\gets$ $\argmin_{\vc x \in \verts \cup \normals}\lvert \Stab_\grpG(\Call{RepOf}{\vc x, \grpG}) \rvert$
                \label{FindTransformation:M}
            \CommentLineEmpty[1]{Any of the elements with the smallest number of equivalent}
            \CommentLineEmpty[1]{vectors of $F'$ is selected:}
            \State $\vc x$ $\gets$ $\Call{Any}{\argmin_{\vc x \in \setM}\lvert\left\{ \vc y \in \verts' \cup \normals' : \vc y \sim \vc x \right\}\rvert}$
                \label{FindTransformation:x}
            \State $\xrep$ $\gets$ \Call{RepOf}{$\vc x$, $\grpG$}
            \State $g_x^{-1}$ $\gets$ inverse of \Call{TransformOf}{$\vc x$, $\grpG$}
            \State $\YY_x$ $\gets$ $\left\{ \vc y \in \verts' \cup \normals' : \vc y \sim \vc x \right\}$
                \label{FindTransformation:YY}
            \State $\setT_y$ $\gets$ $\left\{ \Call{TransformOf}{\vc y, \grpG} : \vc y \in \YY_x \right\}$
            \If{$|\verts| < |\normals|$}
                    \label{FindTransformation:D-start}
                \State $\defD$ $\gets$ $\verts$
                \State $\defD'$ $\gets$ $\verts'$
            \Else
                \State $\defD$ $\gets$ $\normals$
                \State $\defD'$ $\gets$ $\normals'$
            \EndIf \label{FindTransformation:D-end}
            \ForAll{$g_s \in \Stab_\grpG(\xrep)$}
                    \label{FindTransformation:stab-loop}
                \State $\bar g$ $\gets$ $g_s g_x^{-1}$
                \ForAll{$g_y \in \setT_y$}
                    \State $g$ $\gets$ $g_y \bar g$
                        \label{FindTransformation:g}
                    \If{$g \defD = \defD'$}
                        \State \Return $g$
                    \EndIf
                \EndFor
            \EndFor
            \State \Return NULL
        \EndProcedure
    \end{algorithmic}
\emph{Utility functions:}
\begin{itemize}
	\item \Call{Any}{$\XX$} returns an arbitrary element of $\XX$
	\item \Call{RepOf}{$\vc x$, $\grpG$} returns a representative of $\vc x$,
		assuming that all vectors have been classified under $\grpG$ and an
		arbitrary but fixed choice of class representatives has been made
	\item \Call{TransformOf}{$\vc x$, $\grpG$} returns $g_x \in \grpG$
		such that $\vc x = g_x \Call{RepOf}{\vc x, \grpG}$, again assuming that
		vectors have been classified under $\grpG$ and that (at least) one group
		element taking its representative into $\vc x$ is known
\end{itemize}
\end{alg}

We now describe a memory-efficient alternative, shown in
Alg.~\ref{alg:FindTransformation}.
As in \cite{Pook-Kolb2022Exact},
this method is used after ensuring that $F$ and $F'$ have the same number of
vertices and number of normal vectors, and that the respective sets of vectors
can be ordered such that $\vc x_i \sim \vc y_i$ for all $i$.

The main idea is to fix one vector $\vc x$ of $F$ and then construct all
transformations
\begin{equation}\label{eq:all-possible-transforms}
    \setT_x \defeq \bigcup_{\vc y \in \YY} \setT_{xy}
\end{equation}
taking $\vc x$ into any of the vectors $\vc y \in \YY$,
where $\YY$ denotes the vectors of $F'$.
Clearly, if $F$ and $F'$ are equivalent, say $gF = F'$ for some $g\in\grpG$,
then $g$ takes $\vc x$ into one of the vectors $\vc y$ of $F'$ and thus
$g\in\setT_x$.
Choosing $\vc x$ as the vector with the smallest stabilizer and fewest
equivalent vectors of $F'$, $\setT_x$ will often be very small and can
be checked one by one.
However, even if the smallest stabilizer is large, the elements of $\setT_x$
can be enumerated without holding the full set in memory.

Alg.~\ref{alg:FindTransformation} performs this test as follows.
In lines~\ref{FindTransformation:M} and \ref{FindTransformation:x},
$\vc x$ is chosen as the vector with the smallest stabilizer and, if there are
multiple possibilities, then the one with the smallest number of equivalent vectors of $F'$.
In line~\ref{FindTransformation:YY}, we store the set of these equivalent vectors
as $\YY_x$.
Independently from the choice of $\vc x$, let $\defD$ be the smaller of the
sets of vertices and of normal vectors of $F$
(lines~\ref{FindTransformation:D-start}--\ref{FindTransformation:D-end}).
We choose $\defD'$ analogously for $F'$.
Since the stabilizer is a group, we can use methods in \lib{GAP} to iterate
over all its elements in
line~\ref{FindTransformation:stab-loop},
while holding only one element in memory at any given
time.
For each element $g_s \in \Stab_\grpG(\xrep)$ and each $\vc y \in \YY_x$,
we form the transformation (line~\ref{FindTransformation:g})
\begin{equation}\label{eq:xy-transform}
    g = g_{y} g_s g_x^{-1}
\end{equation}
and evaluate if the two sets
$g \defD$ and $\defD'$
are equal.
If they are, then $F$ is equivalent to $F'$ and $gF = F'$.
If they are unequal for all $g_s \in \Stab_\grpG(\xrep)$ and
all $\vc y \in \YY_x$,
then the two faces are inequivalent under $\grpG$.

\subsection{Constructing the orbit of a vector}
\label{sub:orbit-of-vector}

We use a variation of the
standard orbit enumeration technique as implemented, e.g.,
in \cite{GAPSoftware}.
Alg.~\ref{alg:Orbit} constructs the orbit of
a vector $\vc x$ under a group $\grpG$
and stores the group elements
taking $\vc x$ to the elements in its orbit.
These group elements are needed
in the procedure \proc{TransformOf} in
Alg.~\ref{alg:FindTransformation}.
The result is stored as a dictionary, where each key--value pair consists of
an element $\vc y$ of the orbit as key and one arbitrary transformation matrix
taking $\vc x$ into $\vc y$ as value.
We will call such a dictionary an {\em orbit map.}

For vertices, most of the group elements and, in fact, most of the
vertices themselves are not needed in Alg.~\ref{alg:FindTransformation}.
Since child faces are constructed only for the fixed representative parent faces,
only the vertices of the representative facets can appear.
Our orbit algorithm therefore
selectively stores
only some of the group elements, which is decided
in Alg.~\ref{alg:Orbit} using a {\em condition} function.
This significantly reduces the memory usage
for the large orbits of vertices.
For $\BW$, only the group elements corresponding to
$1\,067\,070$
out of all
$201\,343\,200$ vertices are needed.\footnote{
    Because some vertices appear in both facets, this number will vary
    depending on which facets are chosen as representatives.
}

\newcommand{\vorbmap}{{\em orbit\_map}}
\newcommand{\vorbit}{{\em orbit}}
\newcommand{\vpool}{{\em pool}}
\newcommand{\vnewpool}{{\em new\_pool}}
\newcommand{\vcond}{{\em condition}}
\newcommand{\vgens}{{\em gens}}
\newcommand{\vsgens}{{\em stab\_gens}}

The idea of the standard orbit algorithm is to repeatedly apply the generators
of the group to the initial and the newly constructed vectors until no new vector appears.
This is used in Alg.~\ref{alg:Orbit}, where the {\vpool} and {\vnewpool} variables
keep track of which new vectors have appeared in the last iteration.
In lines~\ref{ConstructOrbit:if-we-should-add}--\ref{ConstructOrbit:add-to-orbit-map},
we conditionally store the vector and its transformation in {\vorbmap}.
If all vectors are known, the new pool remains empty and the termination
condition of the while-loop is satisfied.
When constructing the orbits of vertices, the {\vcond} is chosen to evaluate to {\em true}
only when the vector lies in one of the representative facets.
For relevant vectors, {\vcond} is set to always evaluate to {\em true}.

\begin{alg}
    \caption{
        Construct the orbit of a vector $\vc x$ under a group $\grpG$.
        The group is given as a set {\vgens} of transformation matrices generating the full group.
        For $\BW$ we use
        $\text{\vgens} = \{\mat M_1, \mat M_2\}$,
        where $\mat M_1, \mat M_2$ are given in Sec.~\ref{sec:symmetries}.
        The {\vcond}
        is a boolean function of one vector and
        specifies if the transformation matrix should be stored for the given vector.
        This procedure returns
        a set of all vectors in the orbit as well as an orbit map.
        See the main text for details.
    }\label{alg:Orbit}
    \begin{algorithmic}[1]
        \Procedure{Orbit}{$\vc x$, {\vgens}, \vcond}
            \State {\vorbit} $\gets$ $\left\{ \vc x \right\}$
                \label{ConstructOrbit:known-init}
            \State {\vorbmap} $\gets$ new empty Dictionary
            \State \vorbmap$[\vc x]$ $\gets$ identity matrix
            \State {\vpool} $\gets$ copy of \vorbmap
                \label{ConstructOrbit:pool-init}
            \While{{\vpool} is not empty}
                    \label{ConstructOrbit:while}
                \State {\vnewpool} $\gets$ new empty Dictionary
                    \label{ConstructOrbit:new-pool-init}
                \ForAll{$\vc y \in$ keys of \vpool}
                        \label{ConstructOrbit:forall-y}
                    \State $h$ $\gets$ \vpool$[\vc y]$
                    \ForAll{$g \in$ {\vgens}}
                            \label{ConstructOrbit:forall-g}
                        \State $\vc y'$ $\gets$ $g \vc y$
                            \label{ConstructOrbit:gy}
                        \If{$\vc y' \notin$ \vorbit}
                            \State {\vorbit} $\gets$ \vorbit ${}\cup \{\vc y'\}$
                                \label{ConstructOrbit:add-to-orbit}
                            \State $h'$ $\gets$ $gh$
                            \State \vnewpool$[\vc y']$ $\gets$ $h'$
                                \label{ConstructOrbit:add-to-new-pool}
                            \CommentLineEmpty[1]{See the main text for this if-statement:}
                            \If{\vcond($\vc y'$)} \label{ConstructOrbit:if-we-should-add}
                                \State \vorbmap$[\vc y']$ $\gets$ $h'$
                                    \label{ConstructOrbit:add-to-orbit-map}
                            \EndIf
                        \EndIf
                    \EndFor
                \EndFor
                \State {\vpool} $\gets$ {\vnewpool}
            \EndWhile
            \State \Return \vorbit, \vorbmap
        \EndProcedure
    \end{algorithmic}
\end{alg}

\subsection{Constructing the stabilizer of a vector}
\label{sub:stab-of-vector}

\newcommand{\vorblen}{{\em orbit\_size}}
\newcommand{\vstab}{{\em stab}}
\newcommand{\vstabsize}{{\em stab\_size}}

The third change to the method in \cite{Pook-Kolb2022Exact} is an algorithm
to construct the stabilizer of a vector under a group $\grpG$.
Our method is again inspired by a standard orbit-stabilizer algorithm
such as the one implemented in \cite{GAPSoftware}.
Stabilizers are needed
in line~\ref{FindTransformation:stab-loop} of Alg.~\ref{alg:FindTransformation},
where we iterate over all elements of the stabilizer of one of the
representative vectors.
For $\grpG = \grp$, there are in total $8$ representative vectors listed in
Tab.~\ref{tab:BW16-normals-vertices}.
We previously let \lib{GAP} find the stabilizer of a vector.
With the knowledge about each vector's orbit size, however, we can implement a
more efficient method.

\begin{alg}
    \caption{
        Construct the stabilizer of a vector $\vc x$ in $\grpG$
        whose orbit size is known.
        As in Alg.~\ref{alg:Orbit}, the group $\grpG$ is given as a set
        {\vgens} of generator matrices.
        See the main text for details.
    }\label{alg:Stabilizer}
    \begin{algorithmic}[1]
        \Procedure{Stabilizer}{$\vc x$, \vgens, \vorblen}
            \State $\grpG$ $\gets$ \lib{GAP} group from \vgens
            \State {\vstabsize} $\gets$ $|\grpG|/$\vorblen
            \State {\vsgens} $\gets$ new empty List
            \State {\vstab} $\gets$ \lib{GAP} group containing only $\mat I_{16}$
            \State {\vorbmap} $\gets$ new empty Dictionary
            \State \vorbmap$[\vc x]$ $\gets$ identity matrix
            \ForAll{$g \in \grpG$}
                \State $\vc x'$ $\gets$ $g \vc x$
                    \label{Stabilizer:x-prime}
                \If{$\vc x' \in$ keys of {\vorbmap}}
                        \label{Stabilizer:if-in-orbit-map}
                    \State $g'$ $\gets$ {\vorbmap}$[\vc x']$
                        \label{Stabilizer:g-prime}
                    \State $g_s$ $\gets$ $g^{-1} g'$
                    \If{$g_s \notin$ \vstab}
                        \State append $g_s$ to \vsgens
                            \label{Stabilizer:add-gen}
                        \State {\vstab} $\gets$ \lib{GAP} group from \vsgens
                            \label{Stabilizer:update-stab}
                        \If{$\lvert\text{\vstab}\rvert =$ \vstabsize}
                            \State \Return \vstab
                        \EndIf
                    \EndIf
                \Else
                    \State {\vorbmap}$[\vc x]$ $\gets$ $g$
                        \label{Stabilizer:store}
                \EndIf
            \EndFor
        \EndProcedure
    \end{algorithmic}
\end{alg}

In Alg.~\ref{alg:Stabilizer}, we construct elements of the orbit by applying
different group elements to the vector $\vc x$
(line~\ref{Stabilizer:x-prime}).
Any vector $\vc x'$ that is visited this way is stored together with the
corresponding group
element in an orbit map
(line~\ref{Stabilizer:store}).
Whenever we encounter a vector $\vc x'$ previously found, we retrieve the
stored group element $g'$
(line~\ref{Stabilizer:g-prime}).
Since $g\vc x = g' \vc x$, we have $g^{-1} g' \vc x = \vc x$ and so
$g_s \defeq g^{-1} g'$ is an element of the stabilizer of $\vc x$.
If it is not yet an element of the
subgroup $\text{\vstab} \subseteq \Stab_\grpG(\vc x)$ found thus far,
it is added to the list of group generators
in line~\ref{Stabilizer:add-gen}.
After updating {\vstab} in line~\ref{Stabilizer:update-stab},
we check if it is complete by comparing its size against the known stabilizer
size.

This is made efficient by two facts.
First, due to the ``birthday paradox''
\cite[Section~3]{flajolet1992207}, the first coincidence in
line~\ref{Stabilizer:if-in-orbit-map} occurs on average after
$1 + \sum_{n=1}^N \prod_{i=1}^{n-1}(1-i/N)$
group elements
(see the second unnumbered equation below \cite[Eq.~(12)]{flajolet1992207}),
where
$N$ is the size of the orbit of $\vc x$ under $\grpG$.
For $\grp$, this means that the first element of the stabilizers
of the vectors in Tab.~\ref{tab:BW16-normals-vertices}
is found after about $83$ (for $\vc n_1$ and $\vc v_1$)
to $10\,210$ (for $\vc v_2, \vc v_4, \vc v_5$) iterations.
Second, the stabilizers are often generated by very few group elements.
In the case of $\BW$, the set of all $8$ stabilizers
is found within minutes on a single core,
since each stabilizer can be generated by only two generators.

\section{Conclusions}
\label{sec:conclusions}

In this work, we provide a complete account of the relevant vectors,
vertices, and face classes of the
Voronoi region of the Barnes--Wall lattice $\BW$.
This is used to calculate the exact second moment of $\BW$.
In order to obtain these results, we improve our algorithm
\cite{Pook-Kolb2022Exact}, allowing it to be used with larger symmetry groups
than previously possible.
We believe that our algorithm can be used to analyse the Voronoi regions of
many lattices with known symmetry group, potentially even in dimensions higher
than 16.

Using Monte-Carlo integration, the exact value of the second moment
is numerically verified.
Furthermore, it is shown that the variance of the numerical result can be
approximated with much higher accuracy than
conventionally obtained with the
jackknife estimator.
This may provide significant improvements in numerical second moment estimates
in the future.

\balance 
\bibliographystyle{IEEEtran}
\bibliography{references}

\begin{thebibliography}{10}
\providecommand{\url}[1]{#1}
\csname url@samestyle\endcsname
\providecommand{\newblock}{\relax}
\providecommand{\bibinfo}[2]{#2}
\providecommand{\BIBentrySTDinterwordspacing}{\spaceskip=0pt\relax}
\providecommand{\BIBentryALTinterwordstretchfactor}{4}
\providecommand{\BIBentryALTinterwordspacing}{\spaceskip=\fontdimen2\font plus
\BIBentryALTinterwordstretchfactor\fontdimen3\font minus
  \fontdimen4\font\relax}
\providecommand{\BIBforeignlanguage}[2]{{%
\expandafter\ifx\csname l@#1\endcsname\relax
\typeout{** WARNING: IEEEtran.bst: No hyphenation pattern has been}%
\typeout{** loaded for the language `#1'. Using the pattern for}%
\typeout{** the default language instead.}%
\else
\language=\csname l@#1\endcsname
\fi
#2}}
\providecommand{\BIBdecl}{\relax}
\BIBdecl

\bibitem{barnes59}
E.~S. Barnes and G.~E. Wall, ``{Some extreme forms defined in terms of Abelian
  groups},'' \emph{Journal of the Australian Mathematical Society}, vol.~1,
  no.~1, pp. 47--63, Aug. 1959.

\bibitem{forney88pt1}
G.~D. Forney, Jr., ``{Coset codes---part I: Introduction and geometrical
  classification},'' \emph{{IEEE} Trans. Inf. Theory}, vol.~34, no.~5, pp.
  1123--1151, Sept. 1988.

\bibitem{forney88pt2}
------, ``{Coset codes---part II: Binary lattices and related codes},''
  \emph{{IEEE} Trans. Inf. Theory}, vol.~34, no.~5, pp. 1152--1187, Sept. 1988.

\bibitem{hahn90}
A.~J. Hahn, ``{The coset lattices of E. S. Barnes and G. E. Wall},''
  \emph{Journal of the Australian Mathematical Society. Series A}, vol.~49,
  no.~3, pp. 418--433, Dec. 1990.

\bibitem{conway99}
\BIBentryALTinterwordspacing
J.~H. Conway and N.~J.~A. Sloane, \emph{Sphere Packings, Lattices and Groups},
  3rd~ed.\hskip 1em plus 0.5em minus 0.4em\relax New York, NY: Springer, 1999.
  [Online]. Available: \url{https://doi.org/10.1007/978-1-4757-6568-7}
\BIBentrySTDinterwordspacing

\bibitem{adoul95}
J.-P. Adoul, ``Lattice and trellis coded quantizations for efficient coding of
  speech,'' in \emph{Speech Recognition and Coding}, A.~J.~R. Ayuso and
  J.~M.~L. Soler, Eds.\hskip 1em plus 0.5em minus 0.4em\relax Berlin, Germany:
  Springer, 1995, ch.~57, pp. 405--422.

\bibitem{lyu22pke}
\BIBentryALTinterwordspacing
S.~Lyu, L.~Liu, J.~Lai, C.~Ling, and H.~Chen, ``Lattice codes for lattice-based
  {PKE},'' 2022. [Online]. Available: \url{https://arxiv.org/abs/2208.13325}
\BIBentrySTDinterwordspacing

\bibitem{calderbank1998quantum}
A.~R. Calderbank, E.~M. Rains, P.~Shor, and N.~J.~A. Sloane, ``{Quantum error
  correction via codes over GF(4)},'' \emph{{IEEE} Trans. Inf. Theory},
  vol.~44, no.~4, pp. 1369--1387, 1998.

\bibitem{buser1994period}
P.~Buser and P.~Sarnak, ``{On the period matrix of a Riemann surface of large
  genus (with an Appendix by J.H. Conway and N.J.A. Sloane)},''
  \emph{Inventiones mathematicae}, vol. 117, no.~1, pp. 27--56, 1994.

\bibitem{conway82}
\BIBentryALTinterwordspacing
J.~H. Conway and N.~J.~A. Sloane, ``Voronoi regions of lattices, second moments
  of polytopes, and quantization,'' \emph{{IEEE} Trans. Inf. Theory}, vol.
  IT-28, no.~2, pp. 211--226, Mar. 1982. [Online]. Available:
  \url{https://doi.org/10.1109/TIT.1982.1056483}
\BIBentrySTDinterwordspacing

\bibitem{orb4-8-3}
\BIBentryALTinterwordspacing
J.~Mueller, M.~Neunh{\"o}ffer, F.~Noeske, and M.~Horn, ``{orb}, {M}ethods to
  enumerate orbits, {V}ersion 4.8.3,'' Sept. 2019, {GAP package}. [Online].
  Available: \url{https://gap-packages.github.io/orb}
\BIBentrySTDinterwordspacing

\bibitem{ancillary-files}
The arXiv version of this paper has ancillary files containing a catalog of the
  face classes of $\Lambda_{16}$.

\bibitem{Pook-Kolb2022Exact}
\BIBentryALTinterwordspacing
D.~Pook-Kolb, B.~Allen, and E.~Agrell, ``Exact calculation of quantizer
  constants for arbitrary lattices,'' 2022. [Online]. Available:
  \url{https://arxiv.org/abs/2211.01987}
\BIBentrySTDinterwordspacing

\bibitem{macwilliams1977theory}
F.~J. MacWilliams and N.~J.~A. Sloane, \emph{The theory of error correcting
  codes}.\hskip 1em plus 0.5em minus 0.4em\relax Elsevier, 1977, vol.~16.

\bibitem{conway84}
\BIBentryALTinterwordspacing
J.~H. Conway and N.~J.~A. Sloane, ``On the {Voronoi} regions of certain
  lattices,'' \emph{SIAM J. Alg. Disc. Meth.}, vol.~5, no.~3, pp. 294--305,
  Sept. 1984. [Online]. Available: \url{https://doi.org/10.1137/0605031}
\BIBentrySTDinterwordspacing

\bibitem{agrell98}
E.~Agrell and T.~Eriksson, ``Optimization of lattices for quantization,''
  \emph{{IEEE} Trans. Inf. Theory}, vol.~44, no.~5, pp. 1814--1828, Sept. 1998.

\bibitem{Lyu2022better}
\BIBentryALTinterwordspacing
S.~Lyu, Z.~Wang, C.~Ling, and H.~Chen, ``Better lattice quantizers constructed
  from complex integers,'' 2022. [Online]. Available:
  \url{https://arxiv.org/abs/2204.01105}
\BIBentrySTDinterwordspacing

\bibitem{GAPSoftware}
\BIBentryALTinterwordspacing
{The GAP {G}roup}, ``{GAP} {\textendash} {G}roups, {A}lgorithms, and
  {P}rogramming, {V}ersion 4.12dev.'' [Online]. Available:
  \url{https://www.gap-system.org}
\BIBentrySTDinterwordspacing

\bibitem{GAPRepo}
\BIBentryALTinterwordspacing
------, ``{Main development repository for GAP {\textendash} Groups,
  Algorithms, Programming},'' Oct. 2021, {Commit
  401c797476b787e748a3890be4ce95ae4e5d52ae}. [Online]. Available:
  \url{https://github.com/gap-system/gap}
\BIBentrySTDinterwordspacing

\bibitem{gappySoftware}
\BIBentryALTinterwordspacing
E.~M. Bray, ``gappy {\textendash} a {P}ython interface to {GAP}, {V}ersion
  0.1.0a4.'' [Online]. Available: \url{https://github.com/embray/gappy}
\BIBentrySTDinterwordspacing

\bibitem{AEVZ}
E.~{Agrell}, T.~{Eriksson}, A.~{Vardy}, and K.~{Zeger}, ``Closest point search
  in lattices,'' \emph{{IEEE} Trans. Inf. Theory}, vol.~48, no.~8, pp.
  2201--2214, 2002.

\bibitem{allen21}
\BIBentryALTinterwordspacing
B.~Allen and E.~Agrell, ``The optimal lattice quantizer in nine dimensions,''
  \emph{Annalen der Physik}, vol. 533, no.~12, p. 2100259, Dec. 2021. [Online].
  Available: \url{https://doi.org/10.1002/andp.202100259}
\BIBentrySTDinterwordspacing

\bibitem{flajolet1992207}
P.~Flajolet, D.~Gardy, and L.~Thimonier, ``Birthday paradox, coupon collectors,
  caching algorithms and self-organizing search,'' \emph{Discrete Applied
  Mathematics}, vol.~39, no.~3, pp. 207--229, 1992.

\end{thebibliography}

\end{document}